\documentclass[twocolumn,showpacs,floats,floatfix,superscriptaddress,aps,pra]{revtex4-1}
\usepackage{amsfonts}
\usepackage{amssymb}
\usepackage{amsmath}
\usepackage{calc}
\usepackage{graphicx}
\usepackage{bm}

\usepackage[normalem]{ulem}
\usepackage{amsmath,amssymb}
\usepackage{multirow}
\usepackage{xcolor,soul}
\usepackage{srcltx}
\usepackage{hyperref,graphicx}

\def\be{ \begin{equation}}
\def\ee{ \end{equation}}
\def\bea{ \begin{eqnarray}}
\def\eea{ \end{eqnarray}}
\def\bse{ \begin{subequations}}
\def\ese{ \end{subequations}}
\def\bc{ \begin{center}}
\def\ec{ \end{center}}

\begin{document}

\author{Stefano Longhi$^{*}$, Davide Gatti, and Giuseppe Della Valle} 
\affiliation{Dipartimento di Fisica, Politecnico di Milano and Istituto di Fotonica e Nanotecnologie del Consiglio Nazionale delle Ricerche, Piazza L. da Vinci 32, I-20133 Milano, Italy}
\email{stefano.longhi@polimi.it}

\title{Non-Hermitian transparency and one-way transport in low-dimensional lattices by an imaginary gauge field}
  \normalsize


%
\bigskip
\begin{abstract}
\noindent  

Unidirectional and robust transport is generally observed at the edge of  two- or three-dimensional quantum Hall and topological insulator systems. A hallmark of these systems is topological protection, i.e. the existence of propagative edge states that cannot be scattered by imperfections or disorder in the system. A different and less explored form of robust transport arises in non-Hermitian systems in the presence of an {\it imaginary} gauge field. As compared to topologically-protected transport in quantum Hall and topological insulator systems,  robust non-Hermitian transport can be observed in {\it lower} dimensional (i.e. one dimensional) systems. In this work the transport properties of one-dimensional tight-binding lattices with an imaginary gauge field are theoretically investigated, and the physical mechanism underlying robust one-way transport is highlighted. Back scattering is here forbidden because reflected waves are evanescent rather than propagative. Remarkably, the spectral transmission of the non-Hermitian lattice is shown to be mapped into the one of the corresponding Hermitian lattice, i.e. without the gauge field, {\it but} computed in the complex plane. In particular, at large values of the gauge field the spectral transmittance becomes equal to one, even in the presence of disorder or lattice imperfections. This phenomenon can be referred to as {\it one-way non-Hermitian transparency}.  Robust one-way transport can be also realized in a more realistic setting, namely in  heterostructure systems, in which a non-Hermitian disordered lattice is embedded between two homogeneous Hermitian lattices.  Such a double heterostructure realizes asymmetric (non-reciprocal) wave transmission. A physical implementation of non-Hermtian transparency, based on light transport in a chain of optical microring resonators, is suggested.

\end{abstract}



\maketitle

\section{Introduction}

Electrons in two-dimensional structures with a magnetic
field exhibit a wide variety of collective quantum
phenomena such as integer, fractional and quantum spin Hall
effects \cite{r1,r2,r3,r4}. A hallmark of these systems is the
existence of edge states with topological protection that can propagate  
at the boundaries immune of disorder and scattering. 
The robustness of topological states has been actively explored in a wide variety of 
quantum Hall systems, topological insulators, and topological superconductors (see e.g. \cite{r5,r6} and references therein), with potential applications  in metrology, quantum computing and spintronics. 
Recently, a large amount of theoretical and experimental effort has been devoted toward the emulation of quantum Hall behavior in bosonic systems, 
i.e. in ultra-cold gases \cite{r6,r7,r8,r9,r10,r11,r12,r13,r14} and photons \cite{r15,r16,r17,r18,r19,r20}.  Besides of the possibility to explore new quantum phases of matter, in optics 
topological insulation and protection  hold the promise for applications in optical isolation and
robust photon transport.  However,  {\it propagative} states with topological protection do not arise in one-dimensional (1D) systems. A natural question thus arises whether quantum-Hall-like robust transport, which is insensitive to disorder and imperfections, can be realized in 1D systems.  In 1996, Hatano and Nelson \cite{r21} investigated the problem of Anderson localization in a 1D disordered non-Hermitian lattice. They showed that an 'imaginary' (rather than real) magnetic field can prevent Anderson localization, with the appearance of a mobility interval at the center of the band. Such a result, referred to as non-Hermitian delocalization, has been subsequently revisited by several authors \cite{r22,r23,r24,r25,r26,r27,r28,r29,r30,r31} and raised some debate about the nature of eigenstates. Inspired by the Hatano-Nelson delocalization transition, in a recent work \cite{r32} it has been conjectured that robust transport is expected to arise in a rather general class of non-Hermitian 1D lattices, provided that a simple condition for the lattice energy band is satisfied. The proposal of an optical implementation of an 'imaginary' gauge field, based on a chain of coupled optical microrings with tailored gain and loss regions \cite{r32}, renewed the interest in the Hatano-Nelson model.\par
In this paper we present a theoretical study of robust one-way transport in 1D tight-binding lattices with an imaginary gauge field, highlighting the phenomenon of {\it non-Hermitian transparency}: besides  to prevent Anderson localization and restoring mobility, it is shown that in the presence of a sufficiently high imaginary magnetic field structural imperfections and disorder in the lattice become almost one-way {\it transparent}, i.e. they do not scatter back waves and do not substantially alter the spectral transmission  of the lattice.  Such a result follows from the fact that (i) back-scattered waves are {\it evanescent} waves, and (ii) the spectral transmission of the non-Hermitian lattice can be mapped into the one of the corresponding Hermitian lattice, i.e. without the imaginary gauge field, {\it but} computed in the complex energy plane. In particular, at large values of the gauge field the spectral transmittance becomes equal to one, even in the presence of disorder or lattice imperfections, thus realizing non-Hermitian transparency.  Robust one-way transport and transparency effects can be also realized when the imaginary magnetic field is applied to a limited spatial region of the array. In this case,  a suitable double heterostructure system, in which a non-Hermitian disordered lattice is embedded between two homogeneous Hermitian lattices, can realize one-way transparency.

\section{Scattering in a tight-binding lattice  with an imaginary gauge field: robust transport and one-way non-Hermitian transparency}

\subsection{Lattice model}
We consider the hopping motion of a quantum particle in a disordered 1D tight-binding lattice in the presence of an imaginary gauge field \cite{r21}, see Fig.1(a). In the nearest-neighbor approximation, the system is described rather generally by the non-Hermitian Hamiltonian
\begin{eqnarray}
\mathcal{H} & = &  \sum_{n}  \left(   \kappa_{n-1,n} |n-1\rangle \langle n|+  \kappa_{n,n-1} |n
\rangle \langle n-1|\right)  \nonumber \\
&+ & \sum_n V_n |n \rangle \langle n|
\end{eqnarray}
where $|n\rangle$ is a Wannier state localized at site $n$ of the
lattice, $\kappa_{n, n + 1}$ and $\kappa_{n+1, n}$ are the generally asymmetric (i.e $\kappa_{n,n+1} \neq \kappa_{n+1,n}^*$) hopping rates between site $|n \rangle$
and $|n+1 \rangle$, and $V_n$ is the energy of Wannier state $|n
\rangle$. 
After setting $| \psi (t) \rangle = \sum_n c_n (t) | n \rangle$ and assuming $\hbar=1$, the amplitude probabilities $c_n(t)$ satisfy the set of coupled equations
\begin{eqnarray}
i \frac{dc_n}{dt} & = &  \kappa_{n,n+1} c_{n+1}+ \kappa_{n,n-1} c_{n-1}+V_n  c_n \nonumber \\
& \equiv & \sum_m \mathcal{G}_{n,m}c_m
\end{eqnarray}
\begin{figure}[htbp]
  \includegraphics[width=83mm]{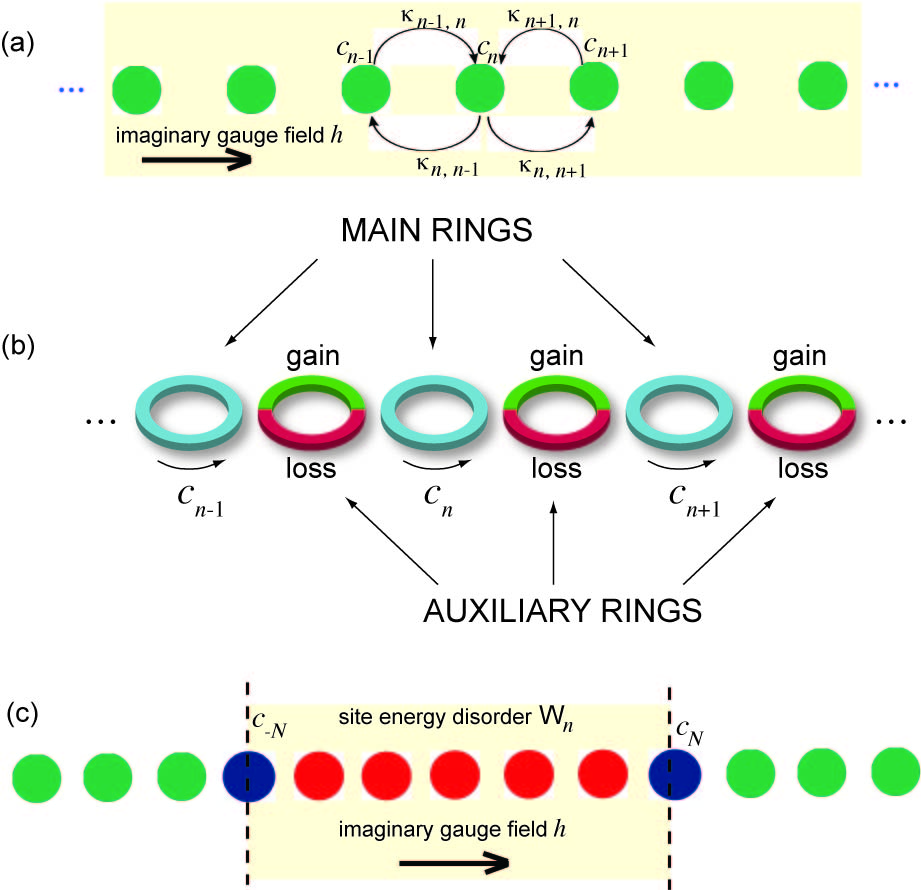}\\
   \caption{(color online) (a) Schematic of a one-dimensional non-Hermitian tight-binding lattice with asymmetric hopping rates induced by an imaginary gauge field $h$ [$\kappa_{n,n+1}= \kappa \exp(-h)$, $\kappa_{n,n-1}= \kappa \exp(h)$]. (b) Optical realization of the non-Hermitian lattice model shown in (a) based on photon hopping in coupled microring resonators. (c) Schematic of a double heterojunction lattice:  a disordered lattice region, with site energy disorder $W_n$  and  applied imaginary gauge field $h$, is connected at sites $n=-N$ and $n=N$ to two Hermitian homogeneous lattices.}
\end{figure}
where we have set 
\begin{equation}
\mathcal{G}_{n,m}=\kappa_{n,n+1} \delta_{n+1,m}+\kappa_{n,n-1} \delta_{n-1,m}+V_n \delta_{n,m}.
\end{equation}
The energy  spectrum $E$ of $\mathcal{H}$ is obtained from the eigenvalues of the tridiagonal matrix  $\mathcal{G}$, i.e. as eigenvalues of the linear equations
\begin{equation}
E \psi_n= \kappa_{n,n+1} \psi_{n+1}+ \kappa_{n,n-1} \psi_{n-1}+ V_n  \psi_n
\end{equation}
with appropriate boundary conditions. 
In the Hatano-Nelson model \cite{r21}, disorder is allowed for the site energies $V_n$, whereas the hopping rates are given by
\begin{equation}
\kappa_{n,n+1}=\kappa \exp(-h) \; ,\;\;  \kappa_{n,n-1}= \kappa \exp(h)
\end{equation}
where $h>0$ describes the effect of the imaginary vector potential and $\kappa$ is the hopping rate in the absence of the imaginary gauge field. 
The Hermitian lattice case is obtained in the limit $h \rightarrow 0$. The possibility to implement asymmetric hopping rates as given by Eq.(5) was originally discussed by Hatano and Nelson for magnetic flux lines in type-II superconductors \cite{r23}. In optics, realization of asymmetric hopping rates was suggested in a few different settings, including engineered coupled waveguide lattices \cite{r33}, active mode-locked lasers with combined amplitude and phase modulators \cite{r33}, and chains of coupled optical microring resonators \cite{r32}. Figure 1(b) shows a possible experimental implementation of an imaginary gauge field $h$ in a chain of coupled microring optical cavities, which has been recently proposed in Ref.\cite{r32}. The resonator chain consists of a sequence of main ring resonators which are indirectly coupled using an interleaved set of auxiliary rings. The auxiliary rings are designed to be antiresonant to the main ring resonators, i.e., the length of
the connecting rings is slightly larger (or smaller) than the main rings so as to acquire an extra $\pi$ phase
shift. To realize a synthetic imaginary gauge field $h$, the auxiliary ring provides amplification in the upper half perimeter, with single-pass amplification
$h$, and balanced loss in the lower half perimeter, with single-pass attenuation $-h$. Light circulation in the main (auxiliary) rings is forced to be counterclockwise (clockwise) by some non-reciprocal element or just because of the excitation conditions. Indicating by
$c_n(t)$ the amplitude of the counterclockwise propagating field in the $n$-th ring in the main resonators,
with a carrier frequency coincident with one longitudinal ring resonance, coupled-mode equations for
the slowly-varying amplitudes $c_n(t)$  can be derived in the mean-field limit  and are precisely of the form (2), with hopping rates given by Eq.(5) \cite{r32}. 
The explicit expression of the hopping rate $\kappa$ in terms of coupling constants of the microrings can be found in Ref.\cite{r32}.
In the optical microring structure of Fig.1(b) the disorder $V_n$ arises because of deviations of microring resonance frequencies from the
reference value due to fabrication imperfections, as well as from slight deviations of the antiresonance condition in the auxiliary
rings \cite{r32}. \par
To properly define the scattering process induced by disorder,  we consider an infinitely extended lattice and assume that disorder or lattice imperfections are confined in the region $ -N \leq n \leq N$ of lattice sites, i.e. we assume 
 \begin{equation}
 V_n=0 \; \; {\rm for} \;\; |n| > N. 
\end{equation} 
 Let us first consider the case of a homogeneous lattice, i.e. $V_n=0$. In this case the (improper) Bloch-Floquet eigenfunctions of $\mathcal{H}$ are given by $\psi_n=\exp(-iqn)$ where $q$ is the Bloch quasi momentum that varies in the first Brillouin zone ($ -\pi \leq q < \pi$). The corresponding eigenenergies $E$ form a tight-binding band with complex energy spectrum described by the dispersion curve 
 \begin{eqnarray}
 E(q) & = & 2 \kappa \cos(q-ih)  \\
 & = & 2 \kappa \cosh(h) \cos(q)+2 i \kappa  \sinh(h) \sin(q). \nonumber
\end{eqnarray}
The group velocity of a wave packet (also related to the current \cite{r21,r23}) , obtained as a superposition of Bloch-Floquet states with carrier wave number $q$, is given by $v_g=-{\rm Re}(dE /dq)=-2 \kappa \cosh (h)  \sin (q)$. Note that a forward-propagating wave ($0<q< \pi$, $v_g>0$) is amplified because ${\rm Im}(E(q))>0$, whereas a backward propagating wave ($-\pi <q< 0$, $v_g<0$) is attenuated because ${\rm Im}(E(q))<0$. As discussed in Ref.\cite{r32}, this circumstance makes wave transport in the lattice highly asymmetric because  backward-propagating waves vanish after some propagation distance due to damping whereas forward-propagating waves are amplified. A different and may be simpler point of view to understand asymmetric (one-way) transport in the lattice when $h \neq 0$ is the following one.  The imaginary gauge field $h$ increases the effective hopping rate in one direction by a factor $\exp(h)$, while in the  opposite direction it is diminished by the factor $\exp(-h)$, i.e. $2h$ represents the exponential suppression of the rates in the two directions. Such an asymmetry in left/right hopping rates leads to a preferred transport in the direction of increased hopping rates, and it is thus similar to asymmetric transport investigated in other physical contexts (see, for example, \cite{referee1,referee2}). Note that, by reversing the sign of $h$, the direction of transparent propagation is reversed.

\begin{figure}[htbp]
  \includegraphics[width=84mm]{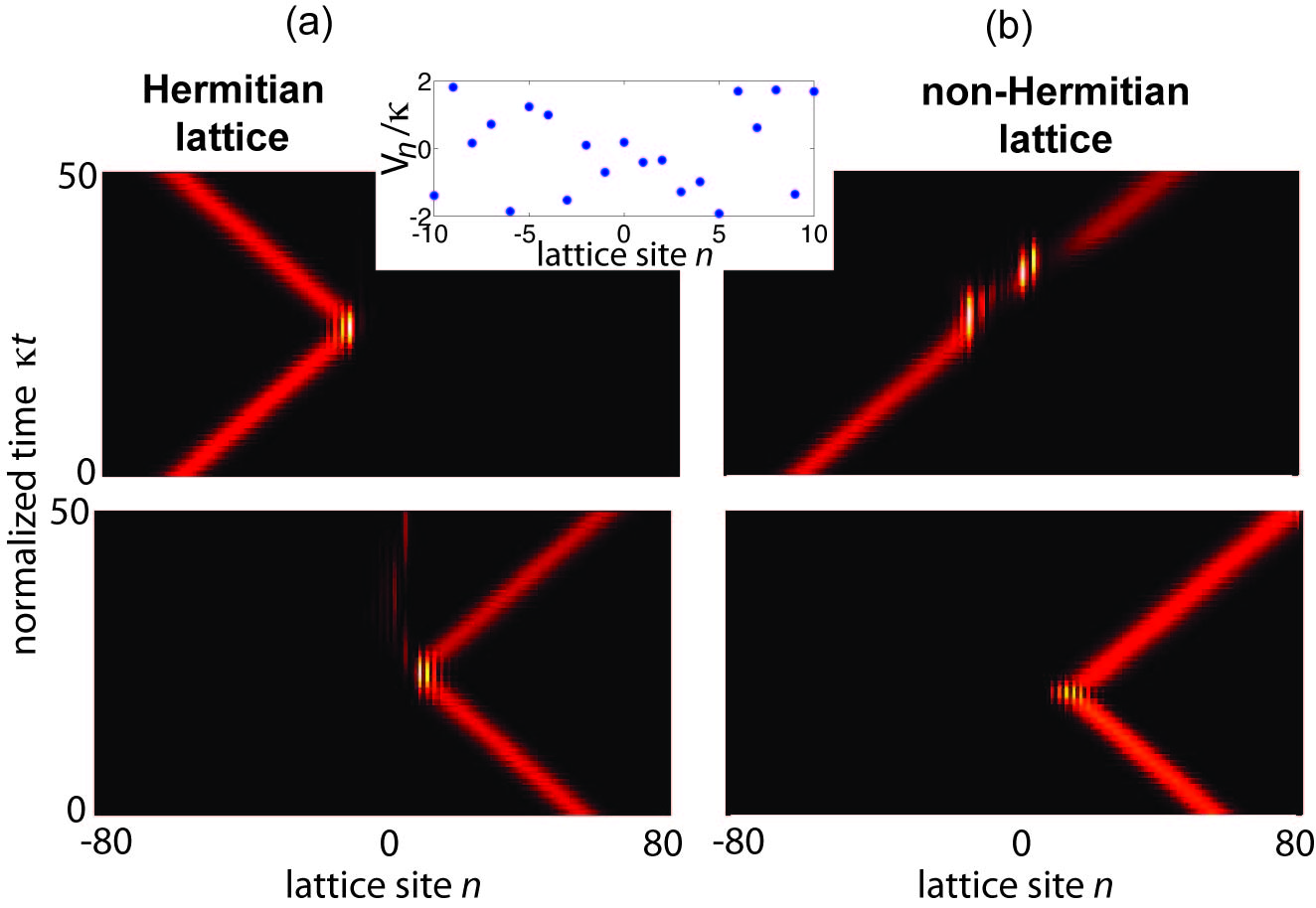}\\
   \caption{(color online) Propagation of a Gaussian wave packet (snapshots of $|c_n(t)|^2 / \sum_n |c_n(t)|^2$) in a uniform lattice with site-energy disorder  for (a) $h=0$ (Hermitian lattice), and (b) $h=0.2$. The energies $V_n$ at sites $-N<n<N$ (with $N=10$) are taken from a uniform distribution  in the range $(-2 \kappa, 2\kappa)$ (see the inset). Upper panels refer to left-side incidence (carrier Bloch wave number of the incident wave packet $q=\pi/2$), whereas lower panels refer to right-side incidence  (carrier Bloch wave number of the incident wave packet $q=-\pi/2$).}
\end{figure}

\subsection{Scattering states and Non-Hermitian transparency}
Extended numerical simulations of wave packet propagation in the Hatano-Nelson lattice showed that transport is robust against disorder and lattice imperfections for forward-propagating waves, but not for backward propagating waves \cite{r32}. An example of this behavior is shown in Fig.2. Such a result was attributed to the asymmetry of the complex energy dispersion curve for forward and backward propagating waves, and was related to the non-Hermitian delocalization transition predicted in the original work by Hatano and Nelson. However, the general properties of scattering states of the lattice and the spectral transmittance for forward waves in the presence of disorder were not investigated in such previous works. Here we close this gap and provide analytical form for the spectral transmission function of forward-propagating waves, predicting an important phenomenon that was not disclosed in previous works: {\it non-Hermitian transparency}. This means that, for a sufficiently large imaginary gauge field, besides to prevent Anderson localization the transport properties in the lattice are completely insensitive to the disorder, i.e. the spectral transmission, both in phase and amplitude, is not altered by the presence of disorder or lattice imperfections.

\begin{figure*}[htbp]
  \includegraphics[width=150mm]{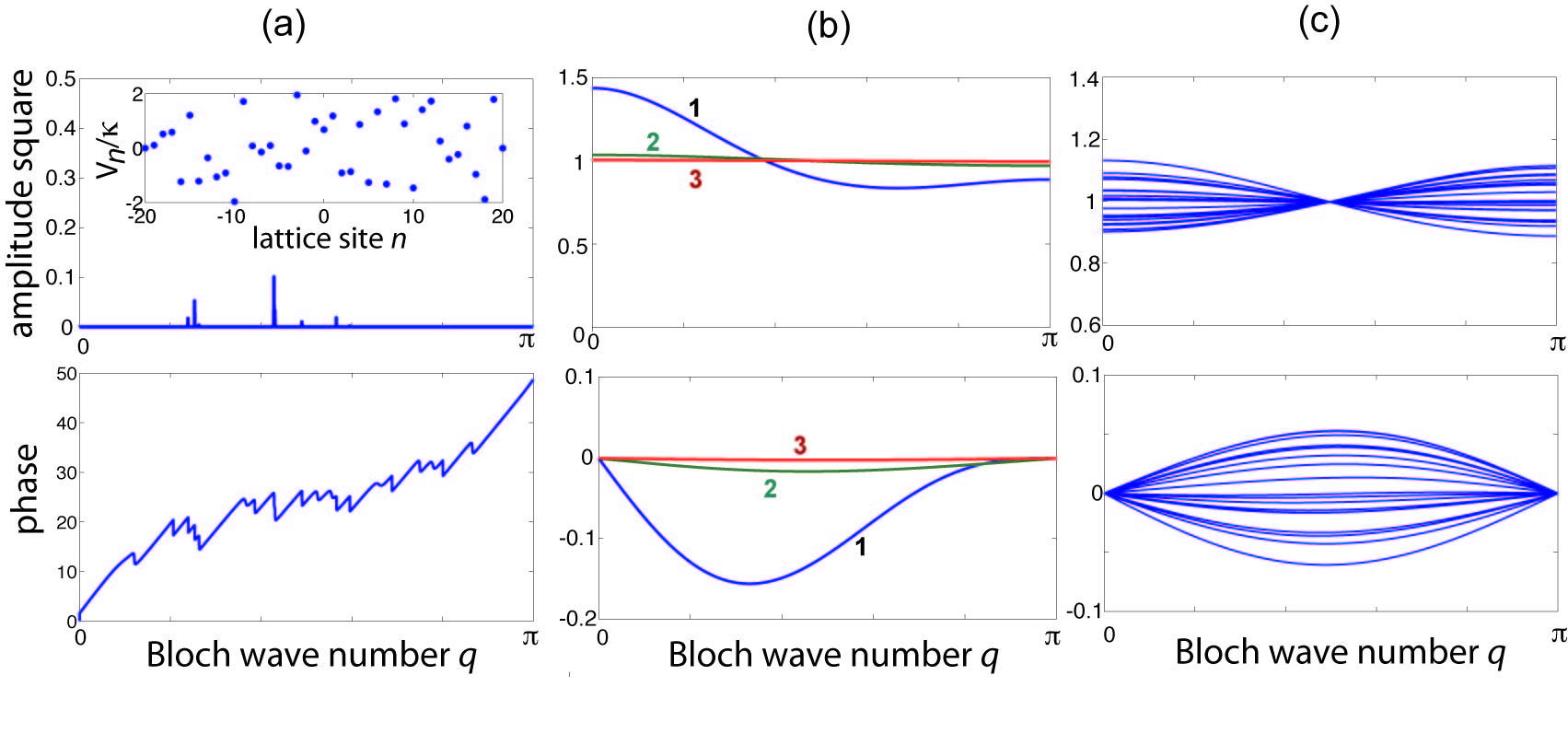}\\
   \caption{(color online) Numerically-computed spectral transmission $t(q)$ (amplitude square and phase) for left-side incidence in a uniform lattice with site-energy disorder at sites $-N \leq n \leq N$ with $N=20$ for (a) $h=0$ (Hermitian lattice) and (b) for a few increasing values of the gauge field $h$ (curve 1: $h=3$, curve 2: $h=5$, curve 3: $h=7$). The energies $V_n$ are taken from a uniform distribution  in the range $(-2 \kappa, 2\kappa)$ [see the inset in (a)]. Panel (c) shows the numerically-computed spectral transmission (amplitude square and phase) for $h=5$ and for 20 different realizations of lattice disorder.}
\end{figure*}

\par
Let us consider a lattice with Hermitian disorder or imperfections of lattice sites $V_n$ confined in the region $-N  \leq n  \leq N$ [see Eq.(6)]. We look for a scattered state solution $\psi$ to Eq.(4) with energy $E=E(q)=2 \kappa \cos(q-ih)$ ($0<q< \pi$), corresponding to a forward propagating wave $\sim \exp(-iqn)$  with Bloch wave number $q$ coming from $n=-\infty$ and incident upon the disordered region of the lattice.   To this aim, let us first notice that in the limiting  case of an Hermitian lattice ($h=0$)  imperfections and disorder generate a backward-propagating wave with opposite Bloch wave number $q$, and the asymptotic form of the scattered state is given by
\begin{eqnarray}
\psi_n= \left\{  
\begin{array}{cc}
\exp(-iqn)+r_0(q) \exp(iqn) & (n \leq -N) \\
t_0(q) \exp(-iqn) & (n \geq N)
\end{array}
\right.
\end{eqnarray}
where $r_0(q)$ and $t_0(q)$ are the spectral reflection and transmission coefficients, respectively. The explicit expression of the spectral transmission and reflection coefficients can be determined by standard transfer matrix method \cite{RR1} and read (see Appendix A for technical details)
\begin{widetext}
\begin{eqnarray}
t_0(q) & = &  \frac{2i \sin (q)}{\mathcal{P}_{11} \exp(iq) -\mathcal{P}_{22} \exp(-iq)+\mathcal{P}_{12}-\mathcal{P}_{21}}  \exp[iq(2N+1)] \\
r_0(q) & = & \frac{\mathcal{Q}_{22}-\mathcal{Q}_{11}  +\mathcal{Q}_{21} \exp(-iq)-\mathcal{Q}_{12} \exp(iq)}{\mathcal{Q}_{11} \exp(iq) -\mathcal{Q}_{22} \exp(-iq)+\mathcal{Q}_{12}-\mathcal{Q}_{21}}  \exp[iq (2N+1)]
\end{eqnarray}
\end{widetext}
where we have set
\begin{eqnarray}
\mathcal{Q}(q) & = & \mathcal{M}_N \times \mathcal{M}_{N-1} \times .... \times \mathcal{M}_{-N} \\
\mathcal{P} (q) & = & \mathcal{M}_{-N} \times \mathcal{M}_{-N+1} \times .... \times \mathcal{M}_{N}
\end{eqnarray}
and
\begin{equation}
\mathcal{M}_n(q)= \left(
\begin{array}{cc}
2 \cos (q)-V_n/ \kappa & -1 \\
1 & 0
\end{array}
\right).
\end{equation}
 In the presence of disorder and for sufficiently large values of $N$, one has $t_0(q) \rightarrow 0$ and $r_0(q) \rightarrow 1$ owing to Anderson localization; see for example Fig.3(a).

 In the presence of the imaginary gauge field ($h \neq 0$) a fully different scenario is found. In fact, owing to energy conservation a scattering process of the incident wave with Bloch wave number $q$ generates a wave with Bloch number $q_1$ satisfying the condition $E(q_1)=E(q)$. Using Eq.(7), such a condition reads explicitly
\begin{equation}
q_1=-q+2ih
\end{equation}
i.e. the scattered wave is an {\it evanescent} wave because it corresponds to a {\it complex} wave number $q_1$. Such a result explains why the imaginary gauge field suppresses back reflections in the lattice. For $h \neq 0$, the scattered state solution to Eq.(4), corresponding to left-side incidence, can be searched in the form [compare with Eq.(8)]
\begin{eqnarray}
\psi_n= \left\{  
\begin{array}{c}
\exp(-iqn)+r (q) \exp[2h(n+N)] \exp(iqn)  \\
 \;\;\;\;\;\;\;\;\ (n \leq -N) \\
t (q) \exp(-iqn) \;\;\; (n \geq N)
\end{array}
\right.
\end{eqnarray}
where $t(q)$, $r(q)$ are the spectral transmission and reflection coefficients. 
A simple relation can be established between $t(q)$, $r(q)$ and their limits $t_0(q)$, $r_0(q)$ in the Hermitian case $h=0$.  In fact, after setting $\psi_n=\phi_n \exp(hn)$ and
$p=q-ih$, from Eqs.(4), (5) and (15) one obtains
\begin{equation}
E \phi_n= \kappa( \phi_{n+1}+  \phi_{n-1})+V_n \phi_n
\end{equation}
with $E=2 \kappa \cos (p)$ and with the asymptotic behavior
\begin{eqnarray}
\phi_n= \left\{  
\begin{array}{c}
\exp(-ipn)+r (q) \exp(2hN) \exp(-ipn)  \\
 \;\;\;\;\;\;\;\;\ (n \leq -N) \\
t (q) \exp(-ipn) \;\;\; (n \geq N)
\end{array}
\right.
\end{eqnarray}
Note that the scattering problem defined by Eqs.(16) and (17) is equivalent to the one of the Hermitian lattice, {\it but} with the real Bloch wave number $q$ replaced by $p=q-ih$. Such an equivalence also follows from an inspection of the band dispersion relation (7), in which we realize that the lattice band of the non-Hermitian system (i.e. for a non-vanishing imaginary gauge field $h \neq 0$) is obtained from the Hermitian one (i.e. in the limit $h=0$) after complexification of the Bloch wave number according to the relation $p=q-ih$.
A comparison of Eqs.(8) and (17) then yields
\begin{eqnarray}
t(q) & = & t_0(q-ih) \\
r(q) & = & \exp(-2Nh) r_0(q-ih).
\end{eqnarray}
It is worth considering the behavior of $t(q)$ and $r(q)$ in the limit of a large imaginary gauge field, i.e. for $h \rightarrow \infty$. Using the expressions (9) and (10) for $t_0(q)$, $r_0(q)$ and from Eqs.(18) and (19) it can be proven that the following asymptotic relations hold (see Appendix B)
\begin{equation}
t(q) \rightarrow 1 \; ,\;\; r(q) \rightarrow 0 \; \; {\rm as} \; \; h \rightarrow \infty. 
\end{equation}
Equation (20) shows that, for a high enough imaginary gauge field $h$, forward wave packet propagation is not affected by disorder or imperfections in the lattice, i.e. {\it unidirectional non-Hermitian transparency} is realized. This phenomenon is illustrated in Figs.3(b) and (c). Figure 3(b) shows the behavior of the spectral transmission $t(q)$ (amplitude and phase) in the same disordered lattice for increasing values of $h$. Note that, according to the asymptotic analysis, as the magnetic flux $h$ is increased the lattice becomes transparent and the spectral transmission $t$ gets close to one. Figure 3(c) shows the behavior of the spectral transmission for a fixed value of the magnetic field ($h=5$) and for 20 different realizations of disorder. The figure clearly indicates that the transmission is almost insensitive to the disorder realization in the lattice.  It should be emphasized that the appearance of non-Hermitian transparency is closely related to the complexification of the Bloch wave number.
In fact, Eq.(18) basically states that the transmission coefficient of propagative waves in the non-Hermitian case is obtained from the expression of the transmission coefficient $t(p)$ in the Hermitian case after the complexification of the Bloch wave number $p=q-ih$. Why is it important such a complexification to realize non-Hermitian transparency? The answer is as follows.  For a real $p$, i.e. for $h=0$, $t(q)$ is the transmission  coefficient of the Hermitian lattice, which is strongly sensitive to the site energy potentials $V_n$: backscattering is fully in action, leading e.g. to Anderson localization in the disordered lattice. When we extend $t(p)$ into the complex plane, $t(p)$ becomes less sensitive to the site potentials $V_n$, and full transparency $t(p) \simeq 1$ is realized in the limit $h \rightarrow \infty$, as briefly sketched in Appendix B. Physically, the lack of sensitivity to $V_n$ for a large imaginary gauge field $h$ relies on the fact that the back-scattered waves are evanescent ones and the coupling to them is weaker and weaker as $h$ is increased: basically in the forward-propagation direction a wave does not see $V_n$ as scattering centers. Such a result holds not only for a random distribution of site energies $V_n$, but also for structural imperfections or defects in the lattice. For example, when $V_n$ is non-vanishing in two sites of the lattice, in the Hermitian lattice ($h=0$) a propagating wave packet undergoes multiple reflections back
and forth between the two defects, like in a Fabry-Perot cavity. This yields multiple transmitted and reflected wave packets, i.e. echoes of the original wave packet are observed (see for example Fig. 3(a) in Ref.\cite{r32}). Application of the imaginary
gauge field to the lattice ($h \neq 0$) makes the two defects invisible and echo effects are suppressed for forward-propagating waves.\\  
A fully opposite behavior is found when considering a wave that propagates in the backward direction. As shown in Fig.2(b), here strong back scattering induced by disorder arises, with the reflected wave packet being amplified. For the ideal case of an imaginary gauge field uniformly applied all along the entire lattice, the scattering problem for right-side incidence  is formally ill posed. The reason thereof is that the scattered states for right-side incidence contain reflected evanescent waves which are undamped as $n \rightarrow + \infty$. However, such unbounded states can be superimposed to describe propagation of localized wave packets, which have physical meaning. In such an analysis, the unbounded evanescent waves lead to an effective amplification of the back scattered wave packet, as observed in Fig.2(b). A detailed analysis of such a case is not given here. Indeed, in the more realistic case considered in the next section, with the imaginary gauge field applied to a limited spatial region of the lattice, the scattering problem for plane waves is well posed for both incidence sides.

\begin{figure}[htbp]
  \includegraphics[width=83mm]{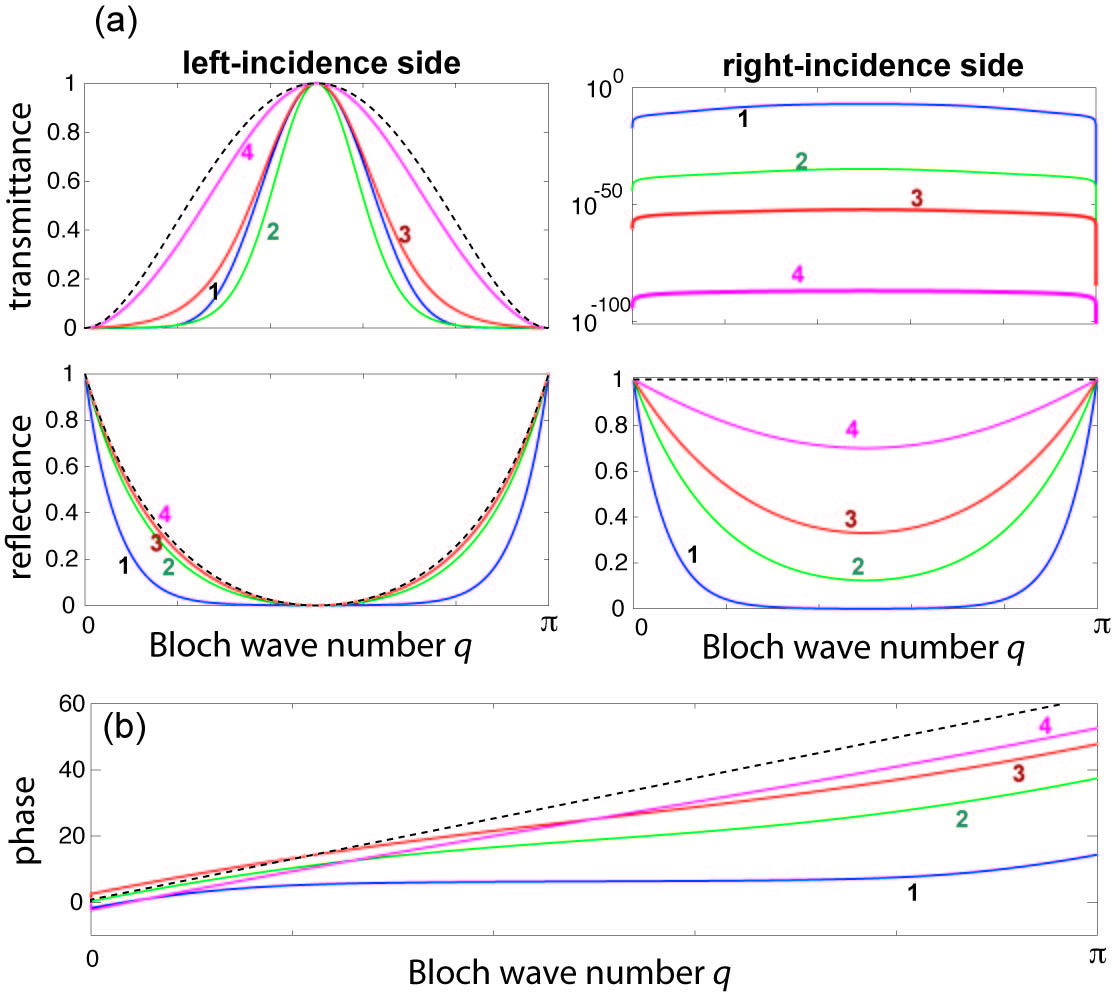}\\
   \caption{(color online) (a) Numerically-computed spectral transmittance ( $|t^{(l,r)}(q)|^2$)  and reflectance ( $|r^{(l,r)}(q)|^2$) of a heterostructure lattice ($N=10$), for left and right sides of incidence, in the absence of disorder. The different curves refer to increasing values of the gauge field $h$. Curve 1: $h=0.2$; curve 2: $h=1$;  curve 3: $h=1.5$;  curve 4: $h=2.5$. The dashed curves show the behaviors of spectral transmittance and reflectance in the $h \rightarrow \infty$ limit [Eqs.(34-37) given in the text]. (b) Behavior of the phase of the spectral transmittance $t^{(l)}(q)$.}
\end{figure}

\section{Double heterostructure lattice and one-way robust transmission}
In the previous section we assumed an infinitely extended lattice with a uniform imaginary gauge field $h$ extended over the entire lattice, both in the ordered and disordered regions. However, in a more realistic system one wishes to apply the imaginary gauge field to a limited spatial region of the lattice, namely to the one with disorder, with the aim to prevent Anderson localization and to suppress the impact of disorder on the spectral transmission. In this case one basically realizes a double heterostructure: a non-Hermitian lattice with disorder embedded between two  Hermitian homogeneous lattices; see Fig.1(c). The Hamiltonian of the heterostructure is described by Eq.(1) with
\begin{eqnarray}
\kappa_{n,n+1} & = & \kappa_{n+1,n}= \kappa  \;\;\;\;\; n \leq -N-1 \; , \; \; n \geq N \nonumber \\
\kappa_{n,n+1} & =  & \kappa \exp(-h) \;\;\;\;\; - N \leq n \leq N-1 \\
\kappa_{n,n-1} & = & \kappa \exp(h)  \;\;\;\;\; - N+1 \leq n \leq N.
\end{eqnarray}
The site energy $V_n$ is assumed uniform and equal to zero in the Hermitian regions ($V_n=0$ for $n < -N$ and $n>N$), whereas in the non-Hermitian region we now assume
\begin{equation}
V_n=-iS +W_n  \; \; \; ( -N<n <N).
\end{equation}
In Eq.(23) $S$ provides an imaginary bias to the energy whereas the real parameters $W_n$ account for disorder and lattice imperfections in the site energies. 
The Hermitian and non-Hermitian sections are attached at the sites $n=-N$ and $n=N$, at which the site energies are  $V_{-N}$ and $V_{N}$ will be determined later.
The spectral transmission and reflection of the structure, for both left ($l$) and right ($r$) incidence sides are defined as the scattering states [Eq.(4)] with the asymptotic behavior
\begin{eqnarray}
\psi_n= \left\{  
\begin{array}{cc}
\exp(-iqn)+r^{(l)}(q) \exp(iqn) & (n \leq -N) \\
t^{(l)}(q) \exp(-iqn) & (n \geq N)
\end{array}
\right.
\end{eqnarray}
for left-side incidence, and 
\begin{eqnarray}
\psi_n= \left\{  
\begin{array}{cc}
\exp(iqn)+r^{(r)}(q) \exp(-iqn) & (n \geq N) \\
t^{(r)}(q) \exp(iqn) & (n \leq -N)
\end{array}
\right.
\end{eqnarray}
for right-side incidence, where $0< q < \pi$ and
\begin{equation}
E=2 \kappa \cos q 
\end{equation}
is the real energy.  
\begin{figure}[htbp]
  \includegraphics[width=83mm]{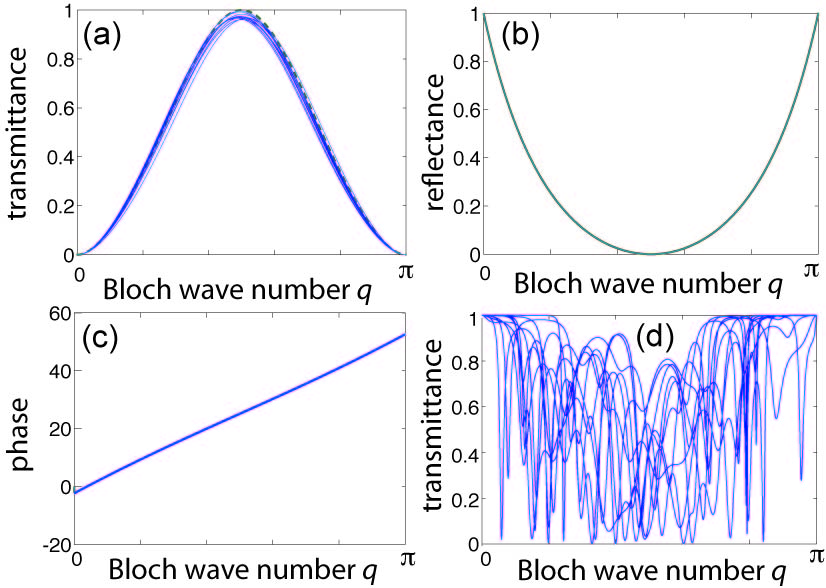}\\
   \caption{(color online) Numerically-computed behavior of (a) spectral transmittance $|t^{(l)}(q)|^2$, (b) reflectance  $|r^{(l)}(q)|^2$, and (c)  phase of $t^{(l)}(t)$ (left side incidence) in a heterojunction lattice with disorder at sites $-N \leq n \leq N$ with $N=10$ for 10 realizations of disorder. The energies $W_n$ are taken from a uniform distribution  in the range $(- \kappa, \kappa)$. The gauge field is $h=2.5$. The dashed curves [almost overlapped with the solid ones in panels (b) and (c)] show the behavior of the spectral functions in the absence of disorder. In (d) the behavior of the spectral transmittance $|t^{(l)}(q)|^2$ in the absence of the junction, i.e for $h=0$, is also shown for comparison.}
\end{figure}
The spectral reflection and transmission coefficients $t^{(l,r)}(q)$, $r^{(l,r)}(q)$ for left and right incident sides  can be determined following the general procedure outlined in the Appendix A. The spectral coefficients are functions of the gauge field $h$ and their values as $h \rightarrow 0$ are denoted by $t_0^{(l,r)}(q)$, $r_0^{(l,r)}(q)$.
 In the absence of the gauge field ($h=0$), the spectral transmission does not depend on the incidence side, i.e. one has
  $t_{0}^{(l)}(q)=t_{0}^{(r)}(q) \equiv t_0(q)$, whereas for $h \neq 0$ a non-reciprocal behavior arises, i.e.  $t^{(l)}(q) \neq t^{(r)}(q)$.
The spectral coefficients are strongly affected by the two interfaces and by the different dispersion relations of the lattices in the regions with and without the applied gauge field. In particular, a major role is played by the complex bias energy $S$ between Hermitian and non-Hermitian regions. The case $S=0$ is a rather simple one to handle analytically, however  regrettably in this case the double heterostructure fails to realize robust and disorder-insensitive transport. In fact, assuming $S=0$ the spectral problem for the amplitudes $\psi_n$  [Eq.(4)] in the $ h \neq 0$ case can be mapped into the one with $h=0$ after the substitution 
\begin{equation}
\psi_n= \phi_n \times \left\{
\begin{array}{cc}
1 & n \leq -N \\
\exp(hn+hN) & -N \leq n \leq N \\
\exp(2hN) & n \geq N ,
\end{array}
\right.
\end{equation}
 i.e. one has
 \begin{equation}
 E \phi_n= \kappa ( \phi_{n+1}+\phi_{n-1}) +V_n \phi_n.
 \end{equation}
 For $S=0$ the following simple relations are thus readily obtained
 \begin{eqnarray}
 r^{(l,r)}(q) & = & r_0^{(l,r)}(q) \nonumber \\
 t^{(l)}(q) & = & \exp(2hN) t_0(q) \\
 t^{(r)}(q) & = & \exp(-2hN) t_0(q). \nonumber
 \end{eqnarray}
  This means that, while the gauge field $h$ makes the transmission non-reciprocal, amplifying forward-propagation waves and attenuating backward-propagating waves, it does not suppress the detrimental effect of disorder as compared to the Hermitian lattice. Such a result seems apparently to contradict the property of non-Hermitian transparency discussed in the previous section.  The reason thereof stems from the fact that non-Hermitian transparency occurs for {\it propagative} waves (i.e. with a real Bloch wave number); however,
 for $S=0$ the incident wave with propagative Bloch wave number $q$ and energy $E(q)=2 \kappa \cos(q)$ coming from the Hermitian lattice section excites, in the non-Hermitian interface region, two {\it evanescent} waves with complex wave numbers $q_{1,2}$ which are
  obtained from the relation (elastic scattering) 
  \begin{equation}
  2 \kappa \cos q= 2 \kappa \cos(q_{1,2}-ih)
  \end{equation} 
i.e. $q_{1,2}= \pm q+ih$. To excite propagative waves in the non-Hermitian interface region and thus to realize non-Hermitian transparency, we introduce a complex energy bias $S$, so that the condition (30) of elastic scattering changes as follows 
  \begin{equation}
  2 \kappa \cos q= 2 \kappa \cos(q_{1,2}-ih)-iS.
  \end{equation} 
For a given value of $q$, we may choose $S$ so as $q_1$ turns out to be real and positive, corresponding to the excitation of a forward-propagating wave in the non-Hermitian interface. 
We typically assume $q= \pi/2$, which corresponds to a propagative wave packet with largest group velocity and minimal distortion \cite{r32}. By imposing $q=q_1= \pi/2$, from Eq.(31) one obtains
\begin{equation}
S=2 \kappa \sinh (h).
\end{equation}
The site energies $V_{\pm N}$ at the interface sites $n=-N$ and $n=N$ are then chosen to obtain zero reflection and full transmission at $q= \pi/2$ in the absence of disorder, i.e. for $W_n=0$ (impedance matching). Their values can be readily calculated by imposing that Eq.(4) is satisfied by $\psi_n= \exp(-iqn)$ over the entire lattice at $q= \pi/2$. Assuming $V_n=-iS$ for $-N<n<N$ and $V_n=0$ for $|n|>N$ one obtains
\begin{equation}
V_{-N}= -i \kappa [1-\exp(-h)] \; , \; \; V_{N}=-i \kappa [\exp(h)-1].
\end{equation}
In the optical implementation of the Hamiltonian (2) shown in Fig.1(b), the imaginary energy bias $S$ as well as $V_{ \pm N}$ just correspond to the introduction of a suitable level of optical loss (absorption or scattering) in the main microrings  at $-N<n<N$. In practice, this can be achieved by a judicious doping of the dielectric material with atomic absorbers or by tailoring the scattering (radiation) losses (i.e. cavity $Q$ factor) of the microrings by e.g. surface roughness control. 
The double heterostructure lattice synthesized in this way realizes one-way transport which is robust against disorder in the non-Hermitian section of the junction. In fact, following a similar perturbative procedure as the one detailed in the Appendix B the asymptotic expressions for the spectral reflection and transmission coefficients, for both left and right incidence sides, can be derived in the limit $h \rightarrow \infty$, which read explicitly
\begin{eqnarray}
t^{(l)}(q) & = & \exp[2iqN-i N \pi] \frac{ 2 i \sin q}{\exp(iq)+i}+O(\epsilon) \;\;\;\; \\
r^{(l)}(q) & = & - \exp(2iqN) \frac{\exp(-iq)+i}{\exp(iq)+i}+O(\epsilon) \\
t^{(r)}(q) & = & O(\epsilon) \\
r^{(r)}(q) & = & 1+O (\epsilon) 
\end{eqnarray}
where $ \epsilon \equiv \exp(-h)$. 
Note that at leading order in $\epsilon$ the spectral coefficients do not depend on the disorder $W_n$. Figure 4(a) shows the numerically-computed behavior of the spectral
transmittance and reflectance for increasing values of the gauge field $h$ in the absence of disorder ($W_n=0$), together with their asymptotic expressions given by Eqs.(34-37).  
The figures clearly shows that the heterostructure is highly non-reciprocal. In particular, at around $q=\pi/2$ transparency for forward-propagating waves, and full rejection of backward waves, is observed. It is also worth commenting on the behavior of the phase $\phi_t$ of the spectral transmission $t^{(l)}(q)$ for forward-propagating waves, which is depicted in Fig.4(b). Note that, as $h$ is increased, the group delay $\tau=-(d \phi_t/dq)$ at around $q=\pi/2$ is non-vanishing and negative, indicating that a wave packet will travel the non-Hermitian region faster as $h$ increases. Such a behavior can be simply explained after observing that the group velocity $v_g$ of the wave packet in the non-Hermitian region of the junction is $v_g= 2 \kappa \cosh h \sin q$, which is $\cosh h $ times larger than the group velocity in the Hermitian regions [see also Fig.6(c) discussed below]. The impact of disorder on transport along the junction is illustrated in Figs.5 and 6. Figure 5 shows the numerically-computed spectral functions for 10 different realizations of disorder in the same heterostructure of Fig.4 for $h=2.5$. Note that the impact of disorder is fully negligible; such a result should be compared with the behavior of the spectral transmittance in the absence of the gauge field, i.e. for $h=0$, which is depicted in  Fig.5(d). The one-way transparency of the heterojunction for a Gaussian wave packet that crosses the disordered region is shown in Fig.6. While for $h=0$ transmission is prevented for both left- and right-side incidence [Fig.6(a)], one-way transmission is observed for $h \neq 0 $ [Figs.6(b) and (c)]. In particular, as $h$ is increased, the wave packet travels faster in the non-Hermitian region [compare Fig.6(b) and (c)], according to the theoretical analysis.

\begin{figure*}[htbp]
  \includegraphics[width=150mm]{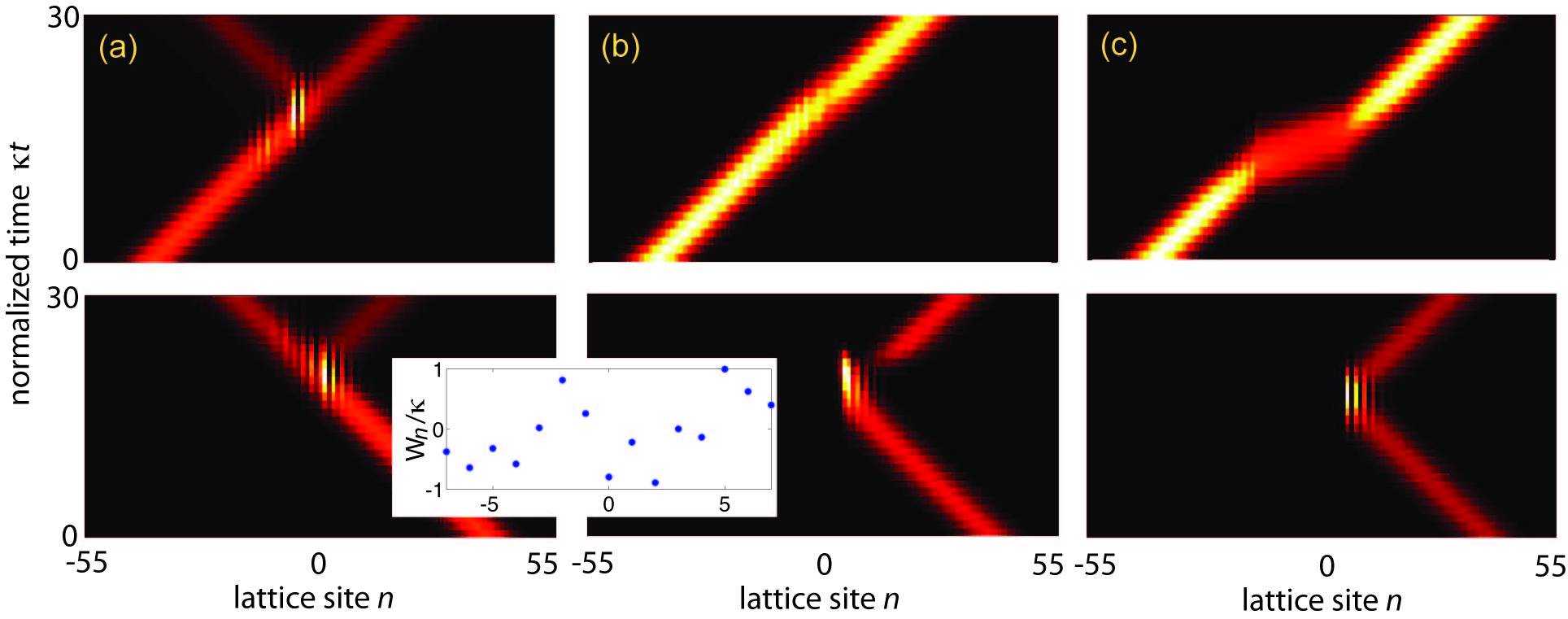}\\
   \caption{(color online) Propagation of a Gaussian wave packet (snapshots of $|c_n(t)|^2 / \sum_n |c_n(t)|^2$) in a heterojunction lattice with disorder  for (a) $h=0$ (Hermitian limit), (b) $h=0.5$, and (c) $h=2$. The energies $W_n$ at sites $-N<n<N$ (with $N=8$) are taken from a uniform distribution  in the range $(-\kappa, \kappa)$ (see the inset). Upper panels refer to left-side incidence (carrier Bloch wave number of the incident wave packet $q=\pi/2$), whereas lower panels refer to right-side incidence  (carrier Bloch wave number of the incident wave packet $q=-\pi/2$).}
\end{figure*}

\section{Conclusions}
Edge states with topological protection that can propagate  
at the boundaries immune of disorder and scattering are known to arise for electrons or bosons in two-dimensional systems subjected to magnetic or synthetic gauge fields. On the other hand, the effect of a magnetic field in one-dimensional systems, such as in a linear tight-binding lattice, is generally trivial and for such a reason it has received few attention so far. Long-range interaction in one-dimensional systems can effectively mimic two-dimensional lattices, leading to fractal energy spectra, flat bands with
localized edge states, and topological many-body states in the presence of synthetic gauge fields \cite{r34}. However, the intrinsic one-dimensional nature of transport is not immune of Anderson localization. In this work we have presented a route toward the realization of robust one-way transport in one-dimensional lattices, based on the application of an {\it imaginary} gauge field. As shown in a pioneering  work by Hatano and Nelson \cite{r21}, in a one-dimensional lattice with disorder application of an 'imaginary' (rather than real) magnetic field can prevent Anderson localization, with the appearance of a mobility interval at the center of the band. 
 In our work we have presented a detailed analytical study of the scattering properties of one-dimensional lattices subjected to an imaginary magnetic field, and have highlighted the phenomenon of {\it one-way non-Hermitian transparency}: besides  to prevent Anderson localization and restoring mobility, we have shown that for a sufficiently high  imaginary magnetic field the structural imperfections and disorder in the lattice become almost {\it one-way transparent}, i.e. they do not scatter back waves and do not substantially alter the spectral transmission  of the lattice for one direction of propagation.  Such a result stems from the fact that back-scattered waves are {\it evanescent} (rather than propagative) waves when an imaginary gauge field is applied. Interestingly, the spectral transmission of the non-Hermitian lattice can be mapped into the one of the corresponding Hermitian lattice, i.e. without the imaginary gauge field, {\it but} computed in the complex energy plane. In particular, owing to complexification of the energy at large values of the gauge field the spectral transmittance becomes equal to one, even in the presence of disorder or lattice imperfections, thus realizing non-Hermitian transparency.  We have also shown that robust one-way transport and one-way transparency can be realized in a more realistic case of a double heterostructure system, in which the magnetic field is applied to a spatially-localized region of the lattice. Our results highlight important and rather unique effects of non-Hermitian transport and are expected to stimulate further theoretical and experimental investigations. In particular, the proposal to implement  imaginary gauge fields in optical systems \cite{r32,r33}, for example using a chain of coupled microrings \cite{r32}, would enable the observation of important phenomena like non-Hermitian delocalization transitions and non-Hermitian transparency in heterojunction lattices.


\appendix

\section{Spectral transmission and reflection coefficients}
In this Appendix we derive explicit forms for spectral transmission and reflection coefficients, corresponding to either left or right incidence sides, in a rather generic lattice described by the Hamiltonian $\mathcal{H}$ given by Eq.(1) in the text. The method is based on a rather standard transfer matrix approach. We assume that the lattice is asymptotically homogeneous and Hermitian, i.e. we will assume 
\begin{eqnarray}
V_n=0 & \;\;\;\;\;\; n<-N, \; \; n>N \\
\kappa_{n,n+1}=\kappa_{n+1,n}=\kappa  & \;\;\;\;\;\;  n<-N, \;\; n \geq N.
\end{eqnarray}
A scattered state solution to Eq.(4) corresponding to left-side incidence has the asymptotic behavior
\begin{eqnarray}
\psi_n= \left\{  
\begin{array}{cc}
\exp(-iqn)+r^{(l)}(q) \exp(iqn) & (n \leq -N) \\
t^{(l)}(q) \exp(-iqn) & (n \geq N)
\end{array}
\right.
\end{eqnarray}
where $t^{(l)}(q)$, $r^{(l)}(q)$ are the spectral transmission and reflection coefficients for left-side incidence, respectively, and $0<q< \pi/2$ is the Bloch wave number of asymptotic plane waves. Similarly, a scattered state solution to Eq.(4) corresponding to right-side incidence has the asymptotic behavior
\begin{eqnarray}
\psi_n= \left\{  
\begin{array}{cc}
t^{(r)}(q) \exp(iqn) & (n \leq -N) \\
\exp(iqn)+r^{(r)}(q) \exp(-iqn) & (n \geq N)
\end{array}
\right.
\end{eqnarray}
where $t^{(r)}(q)$, $r^{(r)}(q)$ are the spectral transmission and reflection coefficients for right-side incidence, respectively. The difference equation (4) can be cast in the following matrix form
 \begin{equation}
 \left( 
 \begin{array}{c}
 \psi_{n+1} \\
 \psi_{n}
\end{array} 
 \right)= \mathcal{M}_n
 \left( 
 \begin{array}{c}
 \psi_{n} \\
 \psi_{n-1}
\end{array}
\right)
\end{equation}
and 
 \begin{equation}
 \left( 
 \begin{array}{c}
 \psi_{n-1} \\
 \psi_{n}
\end{array} 
 \right)= \mathcal{\tilde{M}}_n
 \left( 
 \begin{array}{c}
 \psi_{n} \\
 \psi_{n+1}
\end{array}
\right)
\end{equation}
 where we have set
 \begin{equation}
\mathcal{M}_n=\left( 
 \begin{array}{cc}
 (E-V_n)/ \kappa_{n,n+1} & -\kappa_{n,n-1} / \kappa_{n,n+1} \\
 1 & 0
\end{array}
\right),  
 \end{equation}
 \begin{equation}
\mathcal{\tilde{M}}_n=\left( 
 \begin{array}{cc}
 (E-V_n)/ \kappa_{n,n-1} & -\kappa_{n,n+1} / \kappa_{n,n-1} \\
 1 & 0
\end{array}
\right),  
 \end{equation}
  and $E=2 \kappa \cos q$.
Hence one has
\begin{equation}
 \left( 
 \begin{array}{c}
 \psi_{N+1} \\
 \psi_{N}
\end{array} 
 \right)= \mathcal{Q}
 \left( 
 \begin{array}{c}
 \psi_{-N} \\
 \psi_{-N-1}
\end{array}
\right)
\end{equation} 
 and
 \begin{equation}
 \left( 
 \begin{array}{c}
 \psi_{-N-1} \\
 \psi_{-N}
\end{array} 
 \right)= \mathcal{P}
 \left( 
 \begin{array}{c}
 \psi_{N} \\
 \psi_{N+1}
\end{array}
\right)
\end{equation} 
 where we have set
 \begin{equation}
  \mathcal{Q}=\mathcal{M}_N \times \mathcal{M}_{N-1} \times .... \times \mathcal{M}_{-N+1} \times \mathcal{M}_{-N}.
 \end{equation}
  and
 \begin{equation}
  \mathcal{P}=\mathcal{\tilde{M}}_{-N} \times \mathcal{\tilde{M}}_{-N+1} \times .... \times \mathcal{\tilde{M}}_{N-1} \times \mathcal{\tilde{M}}_{N}.
 \end{equation}
  Note that $\mathcal{\tilde{M}}_n=\mathcal{{M}}_n$ in the absence of the gauge field, i.e. for $h=0$.  
  
From Eqs.(A3), (A4) and (A9) it then follows
\begin{eqnarray}
\left(
\begin{array}{c}
t^{(l)}(q) \exp[-iq(N+1)] \\
t^{(l)}(q) \exp(-iqN) 
\end{array}
\right)=
\left(
\begin{array}{cc}
\mathcal{Q}_{11} & \mathcal{Q}_{12} \\
\mathcal{Q}_{21}  & \mathcal{Q}_{22}
\end{array}
\right) \times \nonumber \\
\left( 
\begin{array}{c}
\exp(iqN)+r^{(l)}(q) \exp(-iqN) \\
\exp[iq(N+1)] +r^{(l)}(q) \exp[-iq(N+1)]
\end{array}
\right) \;\;\;
\end{eqnarray}   
\begin{eqnarray}
\left(
\begin{array}{c}
\exp[iq(N+1)]+r^{(r)}(q) \exp[-iq(N+1)] \\
\exp(iqN)+r^{(r)}(q) \exp(-iqN) 
\end{array}
\right)= \nonumber \\
\left(
\begin{array}{cc}
\mathcal{Q}_{11} & \mathcal{Q}_{12} \\
\mathcal{Q}_{21}  & \mathcal{Q}_{22}
\end{array}
\right) 
\left( 
\begin{array}{c}
t^{(r)}(q) \exp(-iqN) \\
t^{(r)}(q) \exp[-iq(N+1)]
\end{array}
\right) \;\;\;
\end{eqnarray}   
where $\mathcal{Q}_{ik}$ are the elements of the $2 \times 2$ matrix $\mathcal{Q}$. Equations (A13)  and (A14) can be solved for $r^{(l)}(q)$ and $t^{(r)}(q)$, yielding
\begin{widetext}
 \begin{eqnarray}
 r^{(l)}(q)= \exp[iq(2N+1)] \frac{ \mathcal{Q}_{21} \exp(-iq)-\mathcal{Q}_{12} \exp(iq)+\mathcal{Q}_{22}-\mathcal{Q}_{11}}{\mathcal{Q}_{11}\exp(iq)-\mathcal{Q}_{22} \exp(-iq)+\mathcal{Q}_{12}-\mathcal{Q}_{21}} \;\;\;\;\;\;\; \\
 t^{(r)}(q)= \exp[iq(2N+1)] \frac{ 2i \sin q}{\mathcal{Q}_{11}\exp(iq)-\mathcal{Q}_{22} \exp(-iq)+\mathcal{Q}_{12}-\mathcal{Q}_{21}}. \;\;\;\;\;\;\; 
 \end{eqnarray}
 \end{widetext}
 Similarly, from Eqs.(A3), (A4) and (A10) one has 
 \begin{eqnarray}
\left(
\begin{array}{c}
\exp[iq(N+1)]+r^{(l)}(q) \exp[-iq(N+1)] \\
\exp(iqN)+r^{(l)}(q) \exp(-iqN) 
\end{array}
\right)= \nonumber \\
\left(
\begin{array}{cc}
\mathcal{P}_{11} & \mathcal{P}_{12} \\
\mathcal{P}_{21}  & \mathcal{P}_{22}
\end{array}
\right) 
\left( 
\begin{array}{c}
t^{(l)}(q) \exp(-iqN) \\
t^{(l)}(q) \exp[-iq(N+1)]
\end{array}
\right) \;\;\;
\end{eqnarray}   
\begin{eqnarray}
\left(
\begin{array}{c}
t^{(r)}(q) \exp[-iq(N+1)] \\
t^{(r)}(q) \exp(-iqN) 
\end{array}
\right)=
\left(
\begin{array}{cc}
\mathcal{P}_{11} & \mathcal{P}_{12} \\
\mathcal{P}_{21}  & \mathcal{P}_{22}
\end{array}
\right) \times \nonumber \\
\left( 
\begin{array}{c}
\exp(iqN)+r^{(r)}(q) \exp(-iqN) \\
\exp[iq(N+1)] +r^{(r)}(q) \exp[-iq(N+1)]
\end{array}
\right) \;\;\;
\end{eqnarray}   
 which can be solved for $r^{(r)}(q)$, $t^{(l)}(q)$, yielding
 \begin{widetext}
 \begin{eqnarray}
 r^{(r)}(q)= \exp[iq(2N+1)] \frac{ \mathcal{P}_{21} \exp(-iq)-\mathcal{P}_{12} \exp(iq)+\mathcal{P}_{22}-\mathcal{P}_{11}}{\mathcal{P}_{11}\exp(iq)-\mathcal{P}_{22} \exp(-iq)+\mathcal{P}_{12}-\mathcal{P}_{21}} \;\;\;\;\;\;\; \\
 t^{(l)}(q)= \exp[iq(2N+1)] \frac{ 2i \sin q}{\mathcal{P}_{11}\exp(iq)-\mathcal{P}_{22} \exp(-iq)+\mathcal{P}_{12}-\mathcal{P}_{21}}. \;\;\;\;\;\;\; 
 \end{eqnarray}
 \end{widetext}

 \section{Asymptotic behavior of spectral coefficients}
 In this Appendix we prove the asymptotic form of the spectral coefficients $t(q)=t_0(q-ih)$ and $r(q)=\exp(-2Nh) r_0(q-ih)$ given by Eq.(20) in the text. To this aim, let us notice that the expressions of 
 $t_0(p)$ and $r_0(p)$ are given by Eqs.(9) and (10), where $p=q-ih$ and the matrices $\mathcal{Q}$, $\mathcal{P}$ are defined by Eqs.(11-13) given in the text. After setting $\epsilon=\exp(-ip)=\exp(-iq-h)$, we consider the asymptotic forms of $\mathcal{Q}$ and $\mathcal{P}$ as $h \rightarrow \infty$, i.e. $ \epsilon \rightarrow 0$, assuming $V_n/ \kappa$ of order $\sim 1$ (or smaller). Since $\mathcal{P}$ is obtained from the expression of $\mathcal{Q}$ after the substitution $V_n \rightarrow V_{-n}$, we can limit to compute the asymptotic form of $\mathcal{Q}$. To this aim, we note that one has $\mathcal{Q}= \epsilon^{-(2N+1)} \mathcal{R}$ with $\mathcal{R}$ of order $\sim 1$. Owing to the explicit dependence of $t_0(p)$, $r_0(p)$ on $\epsilon=\exp(-ip)$ [see Eqs.(9) and (10)], it is enough to compute the asymptotic expressions of the coefficients $\mathcal{R}_{ik}$ of $\mathcal{R}$ up to the order $\sim \epsilon$ for $(i,j) \neq (1,2)$, whereas the asymptotic expression of $\mathcal{R}_{12}$ should be pushed up to the order $\sim  \epsilon^2$. At such an order of approximation, after some calculations one obtains
 \begin{equation}
 \mathcal{Q}   \simeq \frac{1}{\epsilon^{2N+1}} 
 \left(
 \begin{array}{cc}
1- \epsilon \sum_{l=-N}^N \frac{V_l}{\kappa}  \;\;\;\; - & \epsilon + \epsilon^2 \sum_{l=-N}^{N-1} \frac{V_l}{\kappa}\\
\epsilon & 0
 \end{array}
 \right)
 \end{equation}
 The asymptotic form of the matrix $\mathcal{P}$ is obtained from Eq.(B1)  after the change $V_n \rightarrow V_{-n}$.  Substitution of Eq.(B1) (and a similar expression for $\mathcal{P}$) into Eqs.(9) and (10)
yields
 \begin{eqnarray}
 t(q) & = & t_0(q-ih) =  1+O(\epsilon) \\
 r(q) & = & r_0(q-ih) \exp(-2hN)  \nonumber \\
 & = & \frac{\epsilon V_N}{ \kappa} \exp(2iqN) +O(\epsilon^2)
 \end{eqnarray}
  which prove Eq.(20) given in the text.


\end{document}